\documentclass[12pt,prb]{revtex4}

\usepackage{graphicx}

\newcommand {\bi} {\bibitem}
\newcommand {\be} {\begin{equation}}
\newcommand {\beq} {\begin{eqnarray} \nonumber }
\newcommand {\ee} {\end{equation}}
\newcommand {\eps} {\epsilon}

\newcommand {\epp} {\epsilon^{\prime}}

\newcommand {\Tr} {\mbox{Tr}}

\def \form#1 {eq. (\ref{#1}) }
\def \parziale#1#2  {{\partial {#1} \over \partial {#2}}}
\newcommand{\bq}{\begin{eqnarray}}
\newcommand{\eq}{\end{eqnarray}}

\newcommand{\bc}{\begin{center}}
\newcommand{\ec}{\end{center}}

\newcommand {\bb}{\beta}

\def\(({\left(}
\def\)){\right)}
\def\[[{\left[}
\def\]]{\right]}
\def\bi{\bibitem}

\def \(({\left(}
\def \)){\right)}
\def \[[{\left[}
\def \]]{\right]}

\begin{document}

\title{The liquid-glass transition of silica}
\author{Barbara Coluzzi}
\affiliation{
Dipartimento di Fisica, Universit\`a di Roma 
{\em La  Sapienza},\\ 
INFN Sezione di Roma I, 
Piazzale Aldo Moro, Rome 00185 (Italy)
}
\author{Paolo Verrocchio}
\affiliation{
Dipartimento di Fisica and sezione INFM,\\ 
Universit\`a di Trento, Via Sommarive 14, 
I-38100 Povo Trento (Italy)}

\begin{abstract}
We studied the liquid-glass transition of $SiO_2$ by means of 
replica theory, utilizing an effective pair potential which was proved to 
reproduce a few experimental features of silica.
We found a finite critical temperature $T_0$, where the system undergoes
a phase transition related to replica symmetry breaking, 
in a region where experiments do not show any transition.
The possible sources of this discrepancy are discussed.
\end{abstract}

\maketitle

\section{Introduction}


According to the behavior of their transport coefficients,
supercooled liquids were classified in fragile and strong liquids~\cite{An}.
Close to the experimental glass temperature $T_g$~\cite{footnote}, 
where the system is no longer able to equilibrate on experimental time scales,
the viscosity $\eta$ in strong liquids seems to be in a good 
agreement with the Arrhenius law $\exp(A/T)$, while in the 
fragile ones it is more correctly described by the Vogel-Fulcher-Tamman law 
$\exp(A/(T-T_0))$.
As the dramatic slowing down of dynamics prevents to achieve equilibrium
below $T_g$, the critical temperature $T_0$ can only be inferred by 
extrapolating at lower temperatures the available equilibrium data.
However, this fitting procedure is hardly conclusive, as the
viscosity changes of $14$ order of magnitude in a small window of 
temperatures~\cite{RoSo}. 

In such a classification silica ($SiO_2$) should belong to the class of
strong liquid, though deviations from the pure Arrhenius plot above $T_g$ 
have been recently observed~\cite{HeDiRo}, 
supporting speculations about a fragile-strong transition at temperatures 
higher than $T_g$, that for silica is known to be~\cite{An2} $\sim 1446 K$.

The amorphous network observed in supercooled silica 
has been recently shown~\cite{VoKoBi} to arise 
in molecular dynamics
simulation when using the effective two body potential recently 
introduced by van Best, Kramer and Van Santen~\cite{BeKrVa} (BKS).
However,
due to the limitations of numerical simulations, the transport coefficients
of BKS silica were studied only in a regime of temperature well
above $T_g$, hence the equilibrium behavior at low temperature
is still an open issue. Controversial results indeed arose when the available
data were extrapolated at lower temperatures and the
possibility that the BKS potential rather 
implies a fragile behavior cannot ruled out.
As a matter of fact, fitting data with a VFT law  
yields $T_0 = 2525 K$~\cite{VoKoBi}, while the
power law fit $\eta \propto (T-T_{c})^{\delta}$,
motivated by the predictions of mode-coupling theory~\cite{SiGo}, 
yields the dynamical temperature $T_c \simeq 3300 K$~\cite{HoKoBi3,HoKo,HoKo2}.
On the other hand, the Arrhenius behavior seems to be recovered 
taking into account only the few data available at the lowest temperatures,
suggesting the possibility of a fragile-strong transition at $T_c$. 
Furthermore, the study on the configurational entropy $S_c$~\cite{SaScPo2}, 
shows that the Adam-Gibbs relation $\eta \propto 1/T S_c$
is verified over $4$ decades and suggests that the fragile-strong transition 
might be related to a change of the energy landscape properties at 
low temperatures.

Since the reliability of the BKS potential in describing the low temperature
behavior of the viscosity of silica is still far from being established,
we believe it is instructive to apply the replica 
approach~\cite{Mo,MePa,sferesoft,lj} to this case, looking for a 
possible critical temperature $T_0$ from a 
different route, namely as the point where the system undergoes
a thermodynamic transition~\cite{kauzmann} related to a replica symmetry 
breaking~\cite{MePaVi}. 

\section{Correlated liquid phase}

Let us assume that the glassy dynamics is the signature of the partitioning 
of the whole phase space in an exponentially large number ${\cal N}(f)$ 
of free energy minima 
({\em valleys} in the following), whose exploration in the supercooled phase
is still allowed due to {\em hopping} processes between them.  
The entropic contribution coming from the huge number of different valleys
is called complexity:
\begin{equation}
\Sigma = \log {\cal N}(f)/N
\end{equation}
The 'valleys' must be thought as somewhat more complicated objects
\cite{MePa} than the basins of the 'inherent structures'~\cite{St,St2}, 
as their number can be badly different~\cite{BiMo}.

Generally speaking,
a valley is stable against both an infinitesimal displacement of an extensive
number of particles and a finite rearrangement of a non-extensive number 
of them.
It has been argued~\cite{BiKu} that the complexity in a finite 
dimensional system should be thought as a time-dependent quantity.
The thermodynamic approach~\cite{MePa,sferesoft,lj} rather assumes 
that the valleys are metastable states with a macroscopic time-life, whose 
existence is limited to the region below $T_c$,
as suggested by the analogy with generalized (mean field) spin 
glasses~\cite{KiThWo}. 
We shall exploit this assumption in order to make predictions about 
the supercooled phase, the glass temperature and the glass phase.  

The computation of the complexity $\Sigma(f,T)$ in a given thermodynamic
state of the system (hereafter represented by its temperature $T$)
can be settled by using the method proposed by Monasson~\cite{Mo}, which 
amounts to compute the free energy of a system composed by $m$
identical clones (copies) of the original system.

In the liquid phase, 
upon introducing a suitable coupling, two arbitrary clones can be forced to 
stay in the same valley ({\em correlated liquid phase}), 
but when the coupling vanishes they are always found to be in 
different valleys ({\em uncorrelated liquid phase}).
On the other hand, in the glass phase, the clones remain in the same 
valley even after sending the coupling term to zero~\cite{MePa}.  
Generally speaking, one needs to compute the free energy density of the cloned
system in its correlated liquid phase.
The information on the complexity of the cloned system is obtained when one 
computes the analytic continuation of the correlated liquid 
free-energy density $\phi(m)$
to real values of $m$, by means of~\cite{Mo}:
\begin{equation}
\label{complexity}
{m^2 \over T} {\partial (\phi(m,T)) \over 
\partial m} =\Sigma(m,T).
\end{equation}

Given this very general framework, the aim is to compute
the potential $\phi$, depending both on the standard
thermodynamic parameters (temperature, density, chemical potential, etc..) 
and on the parameter $m$, once the pair potential $V(r)$ has been given.
Three different ways to compute the free energy of a replicated system 
were introduced~\cite{MePa,sferesoft,lj}, each relying on a different set of 
approximations. 
The results shown in the following were obtained within a
harmonic approach, which
only requires the pair correlation functions $g(r)$ computed in the 
liquid phase.
The position $x_i^{(a)}$ of the $a$-th clone of the $i$-th particle is written
as $x_i^{(a)}=z_i + u_i^{(a)}$, isolating the centers of mass 
$z_i$. By assuming that the displacements $u_i^{(a)}$ are 'small' in the
correlated liquid phase, the partition function (in $d$ dimensions) is:
\begin{eqnarray}
Z_m = m^{Nd/2} {\sqrt{2\pi}}^{Nd(m-1)} \: Z_{liq}(T/m)
\left< \exp{\left(-\frac{m-1}{2} \: \Tr \log{{\cal H}/T} \right)} 
\right>_{T/m},
\label{FUNZIONEPARTIZIONE}
\end{eqnarray}
where $Z_{liq}(T/m)$ is the partition function of the liquid phase at the
temperature $T/m$ and ${\cal H}$ is the Hessian matrix of the system.
The average $< \dots >$ is taken over the equilibrium configurations
of the variables $\{z_i\}$.
Introducing the quenched approximation~\cite{MePa}:
\begin{equation}
\left< \exp{\left(-\frac{m-1}{2} \: \Tr \log{{\cal H}/T} \right)} 
\right>_{T/m} \sim
\exp{\left(-\frac{m-1}{2} \left< \Tr \log{{\cal H}/T} 
\right>_{T/m} \right)} 
\label{QUENCHED}
\end{equation}
the free energy $\phi(m)$ is:
\begin{equation}
\phi(m) = f_{liq}(T/m) + \frac{T}{2 m} \left[ \left( m-1 \right)
\left< \Tr \log{{\cal H}/T} \right>_{T/m} -d\: (m-1)\ln{2 \pi} - d \: \log{m}\right]
\label{freearmo}
\end{equation}
The computation of $\left< \Tr \log{{\cal H}/T} \right>_{T/m}$
is simplified by treating the fluctuations of the diagonal term in the
Hessian matrix in a perturbative way~\cite{lj}
which is correct in the high density limit, while the off-diagonal term 
can be taken into account non-perturbatively~\cite{MePa,sferesoft,lj}.

Let us briefly recall the scenario and the predictions of the theory:
\begin{itemize}
\item
In the liquid phase the physical quantities
are obtained in the limit $m \to 1$. In that case
formula (\ref{complexity}) simply reads:
\begin{equation} 
\Sigma(T) = S_{liq}(T)-S_{sol}(T),
\label{compliq}
\end{equation}
$S_{liq}$ being the entropy of the liquid and $S_{sol}$ 
the entropy of a disordered harmonic solid.
\item
If a temperature $T_0$ where complexity vanishes actually exists,
there the ergodicity is broken and the system undergoes the
thermodynamic glass transition.
\item
Choosing as a suitable order parameter of the phase 
transition the correlation function between the different clones, 
the transition turns out to be discontinuous.
Surprisingly enough from the thermodynamic point
of view it is a second order phase transition, the heat capacity 
jumping downwards from liquid to crystal-like values.
\item
In the glass phase the eq.~(\ref{complexity}) at $m=1$
yields a negative complexity, which is meaningless. 
Since in the $m-T$ plane one still finds out a critical 
line $m^{*}(T)$ (lesser than $1$) where $\Sigma=0$, the replica approach
takes $\phi(m^*,T)$ as the physical free energy.
\end{itemize}
Finally, it is interesting to note that the replica scenario shares
many features with the old Adam-Gibbs-Di Marzio one~\cite{kauzmann}.
 
\section{Results}

We analytically studied the glass transition of 
$SiO_2$, when the temperature is lowered and the density is kept
fixed at the value $\rho=2.36 g/cm^3$, which is close to the experimental 
density at ambient pressure~\cite{HoKo,SaScPo2}. We studied the correlated 
liquid phase within the harmonic approach sketched in the previous section, 
describing the interactions by means of the BKS 
potential~\cite{BeKrVa,VoKoBi}: 
\begin{equation}
V^{\eps \epp}(r)=
{\textstyle Q_{\eps \epp} \over {\textstyle r}}+
A_{\eps \epp} e^{\textstyle -B_{\eps \epp} \: r}-
{\textstyle C_{\eps \epp} \over {\textstyle r^6}}+v^{\epsilon \epp}_{sr}(r),
\end{equation}
where $Q_{\eps \epp}=q_{\eps} q_{\epp} e^2$ with $q_{Si}=2.4$, $q_{O}=-1.2$,
$e^2=1602.19/(4\pi8.8542)$ eV $\cdot$ \AA, the other parameter values being 
reported in Tab.~(\ref{tabella}). 
Since the potential has no lower bounds when $r \to 0$, we
followed the procedure of numerical simulations
adding a short-ranged $24-6$ Lennard-Jones term,
which does not affect the thermodynamic quantities~\cite{GuGu}.

In order to obtain the $g(r)$ we resorted to standard theory of 
liquids~\cite{Han1}.
Among the others, the so-called Hypernetted Chain Approximation (HNC) turns
out to be a rather reliable approach in describing fluids with 
electrostatic long range interactions such as silica.
In the HNC free energy the infrared divergence due to Coulomb term is
exactly canceled out by the resummation of the divergences
of the virial expansion~\cite{hncsil}. 
Unfortunately, the HNC closure is not powerful enough to study
the correlated liquid phase. As a matter of fact, 
a thermodynamic inconsistency between the two
different routes for computing the compressibility makes 
the integral
\be
\chi_1=\frac{\bb}{\rho} + \bb \int{d^3 r (g(r)-1)}
\ee
to increase far more rapidly than the quantity
\be
\chi_2=\frac{1}{\rho} \: \left(\frac{dP}{d\rho} \right)^{-1},
\ee
when going to low temperatures.
This prevented to study the whole correlated liquid phase, 
since it is not possible to obtain the $g(r)$ in HNC approximation for 
temperatures below $\sim 3200 K$.

A very simple procedure to deal with 
this difficulty is to get the results at very low temperatures by 
extrapolating the ones obtained at the temperatures allowed
within the HNC approach. 
Since such a procedure might appear not satisfactory
enough, we have also chosen to utilize a different approximation 
interpolating between HNC and Mean Spherical Approximation (MSA) 
introduced by Zerah \& Hansen~\cite{ZeHa} (ZH), which
retains the advantages of the HNC approximation
extending its reliability at much lower temperatures, 
in such a way that no extrapolation is required.
It is interesting to note
that the extrapolated HNC results turn out to be very similar
to the ones obtained by means of the ZH approach, as we shall show.
Finally let us point out that, since the short range behavior is ruled by 
the MSA closure while the long range one is still described by the HNC, the 
cancellation of the infrared divergence described above is unaffected by 
the interpolation.

More in detail, we considered the following integral equations for the
$g^{\eps,\epp}(r)$:
\be
\label{ZH}
c^{\eps} c^{\epp} \rho^2 g_{\eps \epp}(r)=
\exp \left(-\beta V^{\eps \epp}_{(R)}(r) \right)
\left (1+ \frac{\exp \left\{ f^{\eps\epp}(r) \left [ w^{\eps \epp} (r)-\beta 
V^{\eps \epp}_{(A)}(r) \right ] \right \} -1}{f^{\eps\epp}(r)} \right )
\ee
where $V^{\eps \epp}_{(R)}(r) \equiv V^{\eps \epp}(r)-V^{\eps \epp}_{min}$ 
for $r \leq r_{min}$ and zero otherwise, while $V^{\eps \epp}_{(A)}(r) \equiv 
V^{\eps \epp}(r)-V^{\eps \epp}_{(R)}(r)$. 
Moreover 
$w^{\eps \epp}(r) \equiv \int dr^{\prime} h^{\eps \epp}(r-r^{\prime})
c^{\eps \epp}(r^{\prime})$,
where $c^{\eps \epp}(r)$ is the direct correlation function and
$h^{\eps \epp}(r)=g^{\eps \epp}(r)-1$.
The interpolation is achieved by means of the function $f(r)$ 
(here the same for all the components). In the regions where 
$f(r) \sim 1$
the eq.~(\ref{ZH}) gives the well-known HNC equations, while the MSA
ones are recovered for $f(r) \sim 0$.
As we made the choice $f(r)=1-\exp(-r/\alpha)$ the interpolation depends 
only on the parameter $\alpha$.  
Its value can be 
suitably chosen in order to strongly reduce the difference between 
the two compressibilities. 

We numerically solved the equations obtaining the pair correlation 
functions in two different cases. In the former 
we solved the HNC equations, 
corresponding to the $\alpha \to 0$ limit in eq.~(\ref{ZH}), 
while in the latter $\alpha$ has been chosen in such a way 
to reduce the difference $\chi_1-\chi_2$. More precisely, 
we let $\alpha$ to take the constant value $0.7$, which has been checked to 
be the average value minimizing the difference between the two 
compressibilities down to $\sim 4000 K$.
This solution of the integral equations in the $\alpha=0.7$ case gave
our $g_{\eps \epp}(r)$ in the ZH approximation. 

In order to solve the integral equations we discretized 
them, introducing a cut-off $L$ in the real space and a mesh size $a$. 
More in detail, we are showing results obtained when the cut-off in the
real space has the value $L=32$ \AA, considering respectively $4096$ points
(HNC) and $2048$ points (ZH), checking that they do not change
significantly below $T_c$ when $L$ is doubled, though we still observe 
small scale oscillations on the correlation functions depending on the mesh 
size.

\begin{figure}[h!]
\includegraphics[angle=270,width=8.5cm]{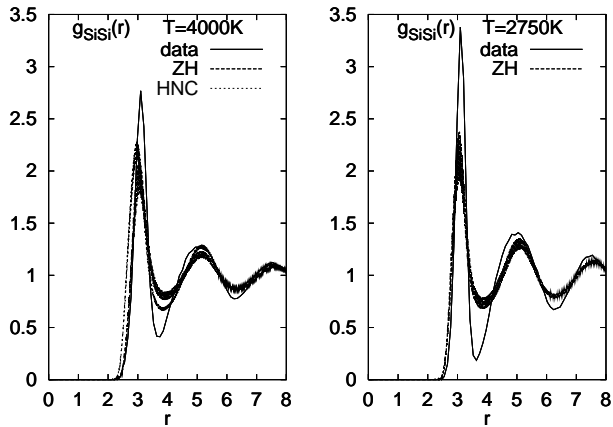}
\caption{\label{gsisi}The analytical $Si-Si$ pair correlation functions 
compared with numerical data by Horbach and Kob.}
\end{figure}

\begin{figure}[h!]
\includegraphics[angle=270,width=8.5cm]{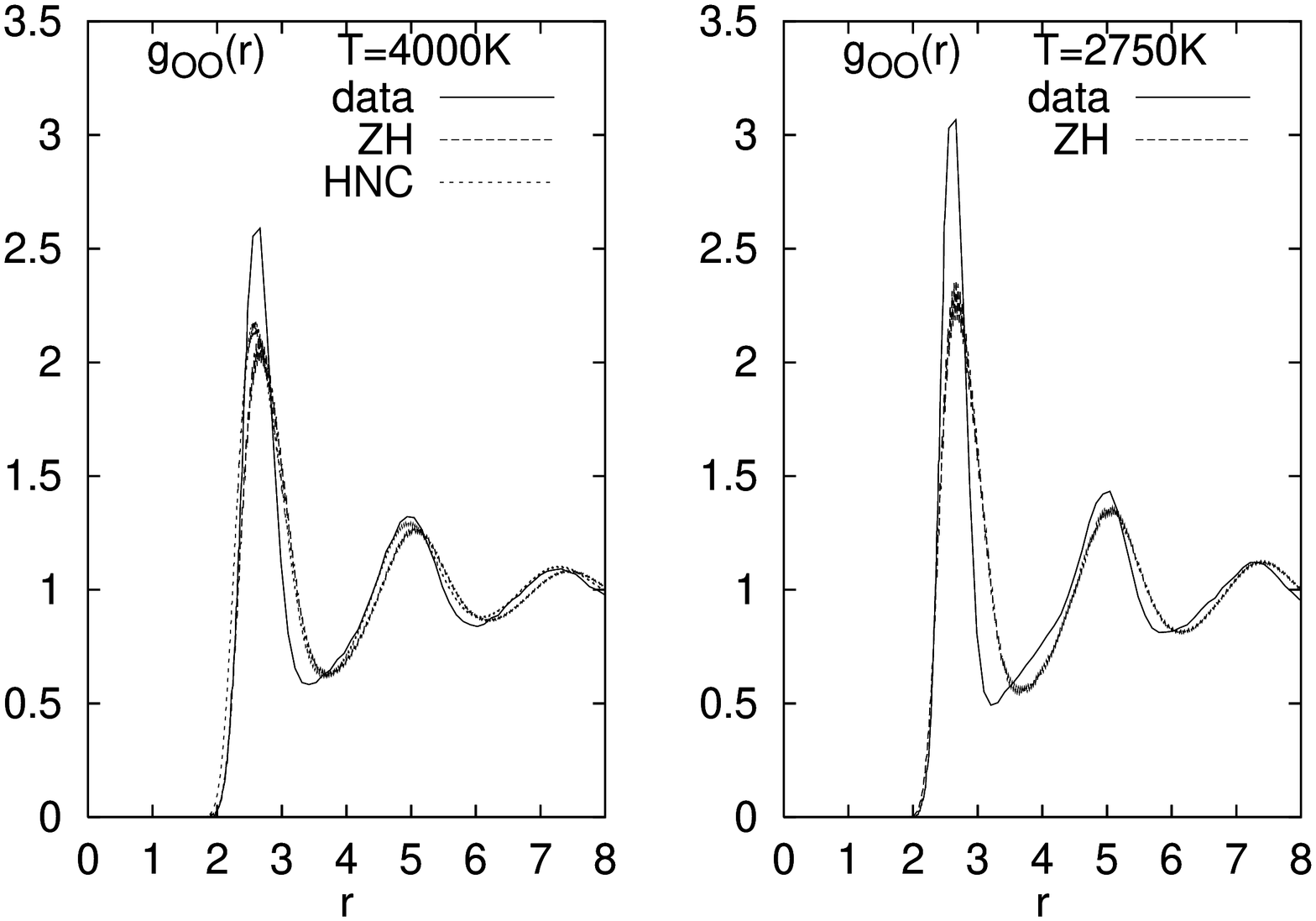}
\caption{\label{goo}The analytical $O-O$ pair correlation functions 
compared with numerical data by Horbach and Kob.}
\end{figure}

\begin{figure}[h!]
\includegraphics[angle=270,width=8.5cm]{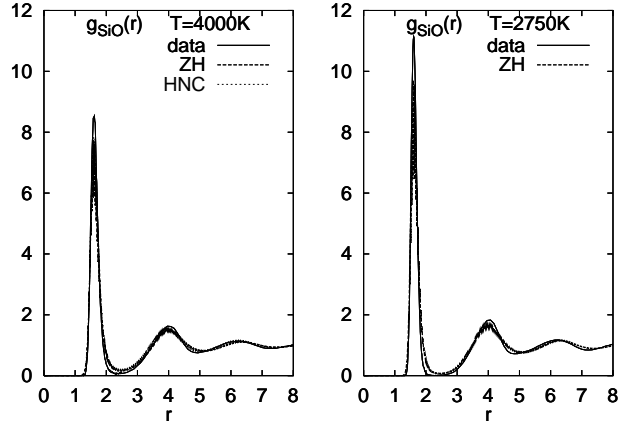}
\caption{\label{gsio}The analytical $Si-O$ pair correlation functions 
compared with numerical data by Horbach and Kob.}
\end{figure}

In figs.(\ref{gsisi},\ref{goo},\ref{gsio}) we show the pair correlation
obtained in the two different approximations, comparing them at
two different temperatures ($4000 K$, $2750 K$) 
with numerical data previously published by Horbach and Kob~\cite{HoKo}.
Let us note that the positions of maxima and minima are quite 
correctly reproduced but there is an error even larger than $10\%$ which 
increases when lowering the temperature on the values of the first peak 
and of the first minimum,
though they roughly cancel out when considering the integral.
Moreover the obtained $g(r)$ show the already mentioned small scale 
oscillations which are related to finite size effects in the momenta domain. 
Therefore the agreement with simulations seems to be not 
as good as in the previously considered cases~\cite{sferesoft,lj}.


\begin{figure}[h!]
\includegraphics[angle=270,width=8.5cm]{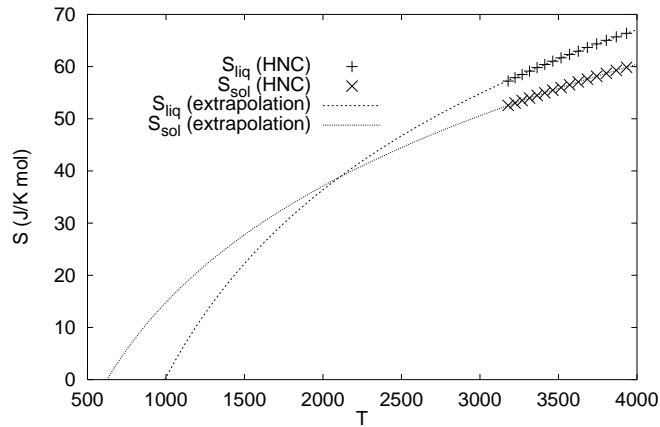}
\caption{\label{enhnc}The liquid and solid entropies as obtained by 
extrapolating HNC results with the functions $a_{liq}+b_{liq}T^{-2/5}$ 
and $a_{sol}+b_{sol} \log T$ respectively. 
Here the thermal contribution and the correct normalization constant
(giving $S=0$ when the phase space volume is equal to $h^3$) are added   
to both of the entropies.}
\end{figure}

As stated before, the HNC results at very low temperatures can only be inferred
by extrapolating the high temperature ones.
That procedure is simplified by the theoretical argument given by
Rosenfeld \& Tarazona~\cite{RoTa}
yielding a $T^{-2/5}$ behavior for the entropy of the liquid 
(apart from the thermal contribution).
Interestingly, also in the ZH approach we found a $S_{liq}$ depending 
linearly on $T^{-2/5}$ in the whole low temperatures region.
On the other hand, the temperature dependence of the disordered
solid entropy is well approximated by a logarithmic behavior,
as expected when only vibrations are taken into account.
In fig.(\ref{enhnc}) we show both the entropies obtained by
the HNC solutions and the results of the extrapolation procedure.


\begin{figure}[h!]
\includegraphics[angle=270,width=8.5cm]{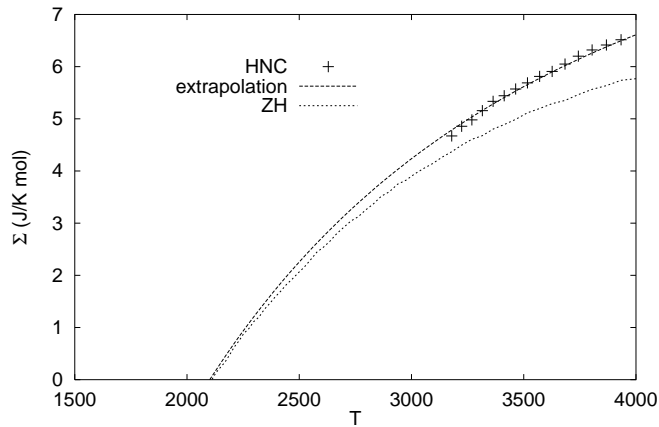}
\caption{\label{comple}The complexity as obtained within the HNC and the 
ZH approximation respectively.}
\end{figure}

The main result of our work is presented in fig.~(\ref{comple}), showing 
the complexity of the system in the liquid phase, computed both in the 
HNC and ZH approaches utilizing the eq.(\ref{compliq}) when $\phi(m)$ is 
given by the eq.~(\ref{freearmo}).
As anticipated, the two curves gives roughly the same $T_0 \sim 2100 K$
yielding a rather sound prediction, which should not strongly depend
on the particular way chosen to describe the liquid phase.

\begin{figure}[h!]
\includegraphics[angle=270,width=8.5cm]{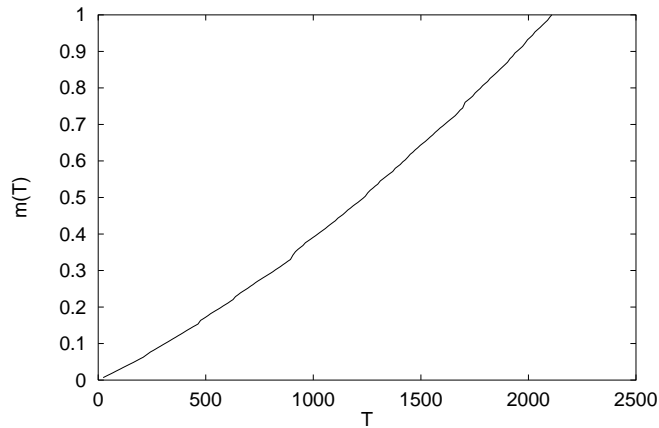}
\caption{\label{emme} The behavior of $m$ as function of the temperature 
in the glassy phase.}
\end{figure}

In the glass phase we want to focus on the predicted behavior
of the thermodynamics parameter $m^*$ as a function of the temperature,
shown in fig.(\ref{emme}).
As explained before, $m^*$ has been found by maximizing the 
free energy~(\ref{freearmo}) in the correlated liquid phase 
with respect to $m$. 
This thermodynamic parameter has a linear behavior, 
shared by all the glassy systems to our knowledge, ranging from fragile 
glasses~\cite{sferesoft,lj} to generalized spin glasses~\cite{CrHoSo},
implying that the quantity $T/m^*$ remains nearly constant in the whole 
glassy phase. 
The analogy with generalized spin-glasses suggests that
$m^*$ might be related to the relative height of the two-peaks equilibrium 
probability distribution of distances between replicas. 
Some numerical evidence for this quantity being non trivial was 
obtained~\cite{distances}.

\section{Discussion}

The study of supercooled BKS silica, by means of the replica
approach dealing with structural glasses introduced by M\'ezard and 
Parisi~\cite{MePa}, which utilizes simple liquid theory in 
order to obtain the pair correlation functions, yielded a liquid-glass 
phase transition at a temperature $T_0 \sim 2100 K$.
This result is quite odd, since it contradicts the
experimental result implying no thermodynamical transition down
to the experimental glass temperature $T_g =1446 K$, 
hence a few comments are strongly required.

As it should be clear from the previous discussion on the behavior of 
viscosity in the BKS silica, our theoretical findings are not ruled out on 
numerical grounds~\cite{HoKo,SaScPo2}, as MD simulations have 
not explored the region where temperatures are much lower than the 
Mode-Coupling one.
Because of that a nearly-Arrenhius behavior (covering no more than two 
decades) has been observed only taking into account a few low temperature data,
whereas a previous extrapolation of data~\cite{VoKoBi} was compatible 
with a divergence at a finite temperature $T_0$ larger than $T_g$.
Therefore one could argue that the too large critical temperature 
found in this thermodynamic approach
is nothing but an artifact of the BKS potential which is not completely
faithful to the actual low temperature behavior of real silica.

On the other hand, as we have already stressed, the $g(r)$ 
obtained from approximate theory of liquids 
show some difference from the ones obtained by MD simulations
utilizing the BKS potential. We want to recall here that both
the approximations involved (HNC and ZH) satisfy the $T^{-2/5}$
behavior for the entropy of the liquid in the whole 
low temperature phase, while the claimed fragile-strong 
transition~\cite{SaScPo2} at $T_c$ points to the failing of this law.
This discrepancy could be the source of a severe
overestimatation of the value of the actual critical temperature.

Hence we are led to consider $2100 K$ rather as a rough estimate of the 
actual glass temperature. As a matter of fact, experiments cannot rule out 
the possibility of a thermodynamic transition at a temperature
lower than $T_g$.

Furthermore, it is worthwhile to note that the approximations 
involved in the computation of correlated liquid free energy
are quite crude. This could lead to an overestimation 
of the critical temperature as well.

An intriguing issue is the possibility 
to gain an insight into the existence of a finite $T_0$ by means
of measurements of the violation of the 
fluctuation-dissipation relation in off-equilibrium experiments.
It has been suggested the existence in structural glasses
of an off-equilibrium regime, called 'aging', reached by 
a system driven out-of-equilibrium by letting it to suddenly cross the 
transition point.

It has been also argued that in the aging regime the 
fluctuation-dissipation relation could still be utilized to define an 
effective temperature $T_{eff}$ which does not depend on the age of the 
system~\cite{CuKu,CuKuPe}.
On very general grounds, it has been moreover conjectured
the equivalence for disordered systems between the ratio $T/T_{eff}$ and the 
parameter describing the replica symmetry breaking, that at one step level is 
nothing but the function $m^*(T)$ computed in the thermodynamic 
approach~\cite{FrMePaPe}.
Of course, the existence of $T_g$ makes the experimental 
situation much more complicated, as for practical purposes the
system falls out-of-equilibrium when crossing $T_g$. At the moment, 
it is unclear how to link an effective temperature measured between 
$T_g$ and $T_0$ to some purely thermodynamic computation. 

Nevertheless, a few attempts on different systems have been 
already performed~\cite{BeCiLa,GrIs} or are in progress~\cite{MaPaVe},
and we believe that even on silica 
those kind of experiments could be useful to clarify many open points.

\section{Acknowledgments}

It is a pleasure to thank G. Parisi to whom we owe 
a lot. We also acknowledge interesting discussions with L. Angelani, 
W. Kob and F. Sciortino.

\newpage

\begin{center}
\begin{table}
\caption{\label{tabella}Parameters of the BKS potential}
\begin{ruledtabular}
\begin{tabular}{|l|l|l|l|}
& A (eV) & B (\AA$^{-1}$) & C (eV $\cdot$ {\AA}$^{-6}$) \\
\cline{1-4}
SiSi & 0.0 & 0.0 & 0.0 \\ 
\cline{1-4}
OO   & 4.87318 & 2.76 & 175.0 \\ 
\cline{1-4}
SiO  & 18003.7572 & 4.8732 & 133.5381\\ 
\end{tabular}
\end{ruledtabular}
\end{table}
\end{center}

\end{document}